# Sub-acoustic resolution photoacoustic imaging through scattering layers using speckle correlations

Benjamin Keenlyside[1,2,*], Arnon A.B. [1,4,*], Dylan Marques[1,2,4], Ivo Vellekoop[4], James Guggenheim[1,2,3]

[1]Department of Cardiovascular Sciences, University of Birmingham, Birmingham, United Kingdom
[2]Department of Electronic, Electrical and Systems Engineering, University of Birmingham, Birmingham, United Kingdom
[3]Department of Medical Physics and Biomedical Engineering, University College London, London, United Kingdom
[4] Biomedical Photonic Imaging, University of Twente, The Netherlands
*Equal contribution.

**ABSTRACT.**
Optical scattering presents a major obstacle to high-resolution imaging in biological tissue and other turbid media. Conventional photoacoustic imaging can partially overcome this obstacle, enabling imaging of optical absorption in the multiple-scattering regime, but its resolution remains limited by acoustic diffraction. In this work we explore a strategy to overcome this limit by exploiting correlations in the illumination patterns produced by coherent scattered light. Combining controlled speckle translations with photoacoustic signal detection, this method enables the recovery of optical resolution images within acoustically selected regions, while overcoming the strict decorrelation range limitations of other speckle correlation techniques. In proof-of-concept experiments, we demonstrate imaging of objects hidden behind an opaque diffuser at sub-acoustic diffraction limited (≤11 μm) resolution, over a >5 mm² field of view much larger than the effective speckle decorrelation range. These results suggest that speckle-correlation-based photoacoustic imaging may offer a route to high resolution imaging of optical absorption under scattering conditions where conventional optical or photoacoustic techniques are fundamentally limited.

Imaging through scattering media, including biological tissue, remains a fundamental challenge in optical science[2]. The strong scattering of light in such materials can severely limit resolution or prevent imaging entirely.

One method of imaging in the presence of strong optical scattering is photoacoustic tomography (PAT) [3]. PAT relies on the generation of ultrasound waves from absorbing structures illuminated by short pulses of light. In this way, optical absorption contrast is encoded onto the amplitude of an acoustic wave, which is more weakly scattered than light in many materials. This approach both circumvents the multiple scattering that rapidly degrades the resolution of purely optical techniques and provides direct access to rich, label-free endogenous absorption contrast from intrinsic chromophores[3-5]. As a result, PAT has emerged as a powerful modality for imaging endogenous contrast in biological tissue, enabling visualisation of anatomical structures or function at depths inaccessible to conventional high-resolution optical microscopy[2]. However, the spatial resolution of PAT is limited by acoustic diffraction to tens to hundreds of microns.

One method that has been explored to surpass the acoustic diffraction limit is to exploit the optical-wavelength scale structure of 'speckle patterns'[6], formed by the interference of sufficiently coherent light. Previous work has used many random speckle illumination patterns to produce signal fluctuations whose statistics encode high spatial frequency information[7-9]. Photoacoustic speckle fluctuation techniques of this kind have been able to surpass the acoustic diffraction limit, but resolution improvements remain relatively modest (~1.4 – 2× smaller than the acoustic diffraction limit).

In this work, we explore a different approach, exploiting correlations present in coherently scattered light under controlled perturbations to the illumination beam, providing additional information to enable images to be reconstructed at resolutions significantly surpassing the acoustic diffraction limit. Specifically, we introduce photoacoustic speckle correlation imaging, a method that uses controlled shifts of speckle illumination to achieve optical-scale resolution imaging of absorbing structures through strongly scattering media.

Our approach builds on methods originally demonstrated in optical speckle-correlation imaging, first introduced by J. Bertolotti et. al. [10], which utilises the angle-invariance of the scattering processes in scattering layers (also known as the 'angular memory effect'[11-13]). Small changes to the illumination angle produce translated, but otherwise nearly identical, speckle patterns, allowing controlled lateral shifts of the illumination.

Translating a speckle pattern over a hidden object in this way causes changes in the overlap between the illumination and the object, modulating the relevant feedback signal. The resulting signal variations can be used to computationally recover the object, despite neither the speckle pattern nor the object being known[14], provided that the angular speckle decorrelation range (or 'angular memory range') exceeds both the scan range and the object's spatial extent. Optical speckle correlation methods using this effect have been shown to be capable of high-resolution imaging of reflective and fluorescent objects hidden behind strong optically scattering layers [10,15,16]. However, they are unable to image optical absorption. Additionally, these methods usually require that the speckle decorrelation range exceed

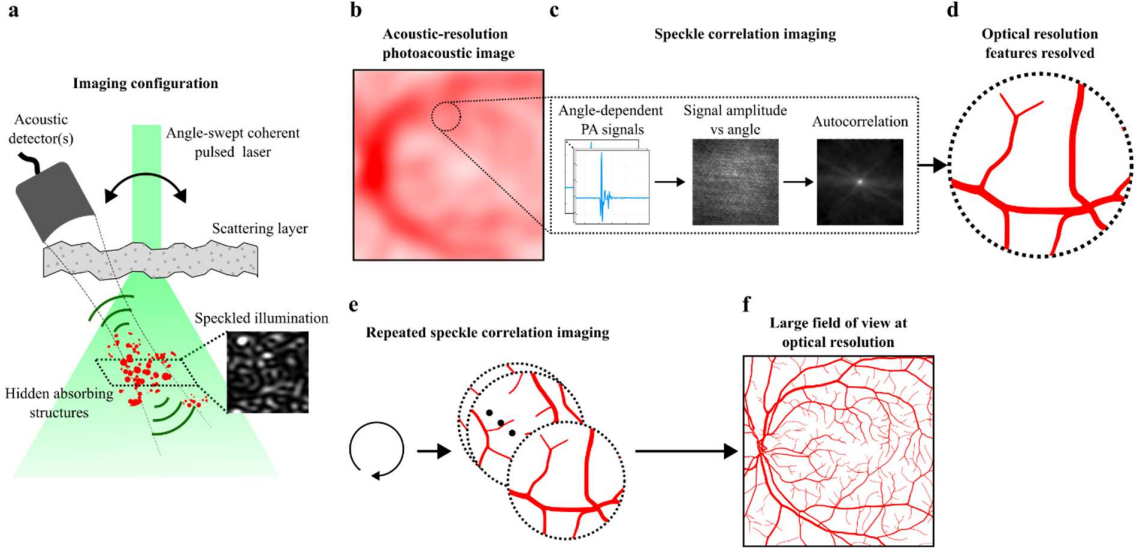

*Figure 1 – Principle of photoacoustic imaging using speckle correlations. (a) Imaging configuration, in which absorbing structures of interest are hidden behind a strongly optically scattering layer. (b) Acoustic-resolution image acquirable through conventional photoacoustic imaging, with a region of interest selected (e.g. using a focused ultrasound detector). (c) Overview of acquisition and reconstruction processes involved in photoacoustic imaging using speckle correlations. Photoacoustic signals are recorded for different illumination angles, and the amplitude of these signals as a function of angle extracted. An image is then reconstructed by first recovering its autocorrelation and then inverting the autocorrelation, yielding (d) an optical resolution image. (e) Repeating this process for several different acoustically selected patches allows (f) an optical resolution image of a larger field of view (beyond the speckle decorrelation range) to be obtained. Vessel masks modified from [1].*

the size of all illuminated objects, and in practice this range is often less than 100µm[17].

The photoacoustic speckle correlation imaging approach proposed here offers the prospect of combining the optical resolution capabilities of speckle correlation techniques with photoacoustic imaging's access to rich optical absorption-based contrast. Simultaneously, the ability to isolate detected photoacoustic signals to an acoustically-selected 'patch' significantly relaxes the usually strict imaging requirements on the speckle decorrelation range, opening the way to imaging of larger structures under a wider range of conditions. Additionally, unlike work relying on methods such as wavefront-corrections[18], this approach can function entirely non-invasively and without lengthy medium-specific calibration.

**Principle**

In the imaging scenario shown in Figure 1a, a hidden object lies behind a strongly scattering layer that prevents conventional high-resolution imaging. Illuminating this object using a short pulse of laser light causes it to emit ultrasound waves due to the photoacoustic effect. These acoustic waves can travel through optically scattering layer without significant distortion and be detected using an acoustic detector. This allows for a photoacoustic image of the object to be obtained at acoustic diffraction-limited resolution, as illustrated in Figure 1b.

To enable imaging at sub-acoustic diffraction-limited resolution, a high-coherence laser is used for photoacoustic excitation, which results in a pseudo-random speckled illumination pattern after light scattering. In this condition, only the regions of the object overlapping with the bright regions of the speckle pattern will produce photoacoustic waves, resulting in acoustic emissions with an amplitude proportional to the overlap between the object and the speckle pattern. When the angle of the excitation beam incident on the scattering layer is scanned, the angle-invariance of the light scattering will result in a translation of the overall speckle pattern without distortion. Scanning the speckle pattern in this way causes changes in the mutual overlap of the pattern and the object, producing corresponding changes in the amplitude of generated photoacoustic waves, which encodes information about the structure of the hidden object.

To reconstruct an image using this information, an autocorrelation-based reconstruction process is used[19], illustrated in Figure 1c. First, the amplitude of photoacoustic wave in each of the recorded signals is estimated, giving angle-resolved amplitude across two orthogonal axes. This gives a measurement of the cross-correlation of the object and speckle pattern, shown in Equation 1:

$$I(\theta, \phi) = \iint O'(x,y) \cdot S(x - \Delta x, y - \Delta y) \, dx \, dy \quad (1)$$
$$= [O' \star S](\theta, \phi)$$

Here, $O'(x,y) = O(x,y) \cdot H(x,y)$ is the effective object seen by the detection system, where $O = \mu_a \hat{\Gamma}$ denotes the object's photoacoustic contrast (with $\mu_a$ being the optical absorption coefficient and $\hat{\Gamma}$ the photoacoustic efficiency), $S(x,y)$ is the speckle intensity distribution, and $H(x,y)$ is the transducer's spatial response function. The quantities $\Delta x$ and $\Delta y$ give the lateral shift of the speckle pattern produced when the illumination beam is tilted by angles $\theta$ and $\phi$. The symbol $\star$ denotes the correlation operator.

The combination here of $O$ and $H$ into $O'$ is what enables the relaxation of the usually strict speckle decorrelation range requirements of speckle correlation techniques. Rather than needing the speckle decorrelation range to be larger than the complete illuminated object $O \cdot S$, we only require this range is larger than the region selected by the transducer response $H$.

Recovering an image of the object from this measurement requires deconvolving the object and speckle pattern. As stated previously, we exploit the properties of optical speckle to make this problem tractable. Specifically, we utilise the property that the autocorrelation of a speckle pattern is a sharply peaked function[10,20]. This means that the autocorrelation of the measured signals is nearly identical to the autocorrelation of the object itself, shown in Equation 2.

$$\begin{aligned} I \star I &= [O' \star S] \star [O' \star S] \\ &= [O' \star O'] \star [S \star S] \\ &\approx [O' \star O'] + C \end{aligned} \quad (2)$$

Reconstructing an image of the hidden object $O'$ can then be achieved by estimating and subtracting background term $C$ followed by inverting the autocorrelation. Inversion of the autocorrelation can be achieved using phase retrieval methods[21,22], and in this work a continuous hybrid input-output phase retrieval algorithm[23,24] was used.

This process is expected to allow image reconstruction at a resolution defined by the size of an optical speckle (approaching $\lambda / 2$ in the limit), inside an acoustically selected patch, illustrated in Figure 1d. To obtain a high-resolution image of a much larger field of view this process could be repeated across a number of acoustic patches, and the recovered images combined via mosaicking, illustrated in Figure 1e-f.

**RESULTS**

As a proof-of-concept investigation of the capabilities of photoacoustic speckle correlation imaging, an experiment was performed involving imaging a simple object hidden behind an optically scattering layer.

The experimental setup is shown in Figure 2. A 220 grit ground glass diffuser was illuminated with a laser beam provided by a 532nm coherent pulsed laser (0.3mJ pulse energy). A MEMS mirror was used to control the angle of incidence of the beam on the diffuser. An absorbing object, consisting of a numerical digit of maximum dimension 74µm plotted onto a film photomask, was hidden approximately 13mm behind the diffuser. The speckle pattern in the object plane had a diameter of approximately 4mm, and a speckle grain size of <5µm. Note that the diameter of the speckle pattern is much larger than range over which it was translated during imaging (<0.5mm). Photoacoustic emissions from the absorbing object were detected using a 25 MHz focused ultrasound transducer with a focal length of 12.5mm and a diameter of 6mm. This transducer was mounted on a pair of translation stages, allowing it to be translated across a plane normal to the optical axis of the system. Complete details of the experimental system used can be found in the supplementary methods section.

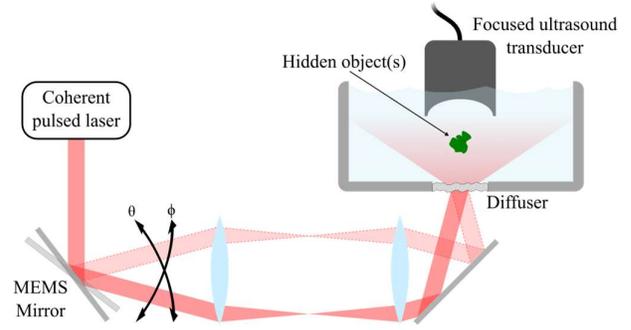

*Figure 2 – A simplified schematic of the experimental setup used to perform photoacoustic speckle correlation imaging. Light from a coherent pulsed laser is used to illuminate a diffuser, with the angle of incidence of the beam controlled by a Micro-Electro-Mechanical System (MEMS) mirror. A focused ultrasound transducer is used to detect the resulting photoacoustic emissions from objects hidden behind the diffuser.*

An acoustically-resolved photoacoustic image of the object was then acquired by raster scanning the transducer and recording time-resolved signals at each position, using the experimental configuration illustrated in Figure 3a. A maximum intensity projection through a 360µm thick x-y slice through the recorded image volume is shown in Figure 3b, with an example time trace in Figure 3c. As expected, the structure of the digit was unresolved, as its size (74 µm) is well below the transducer's resolution (~200 µm FWHM).

To explore sub-acoustic-wavelength spatial resolution imaging, the transducer was returned to the position giving the greatest photoacoustic signal amplitude, placing the hidden object inside the acoustic focus. To sweep the speckle pattern, the angle of the excitation beam was varied along two orthogonal axes using the MEMS mirror (Figure 3d), and an acoustic time series waveform was recorded for each angle pair. The peak amplitude of the photoacoustic signal in each waveform was extracted, and plotted as a function of angle in Figure 3e. The autocorrelation of the variation in photoacoustic signal amplitude with angle was then calculated, shown in Figure 3f. Thresholding was applied to remove the background term and windowing was used to impose a support constraint on the reconstruction. Full details can be found in the supplementary methods.

An image was then reconstructed by inverting the autocorrelation via phase retrieval, shown in Figure 3g.

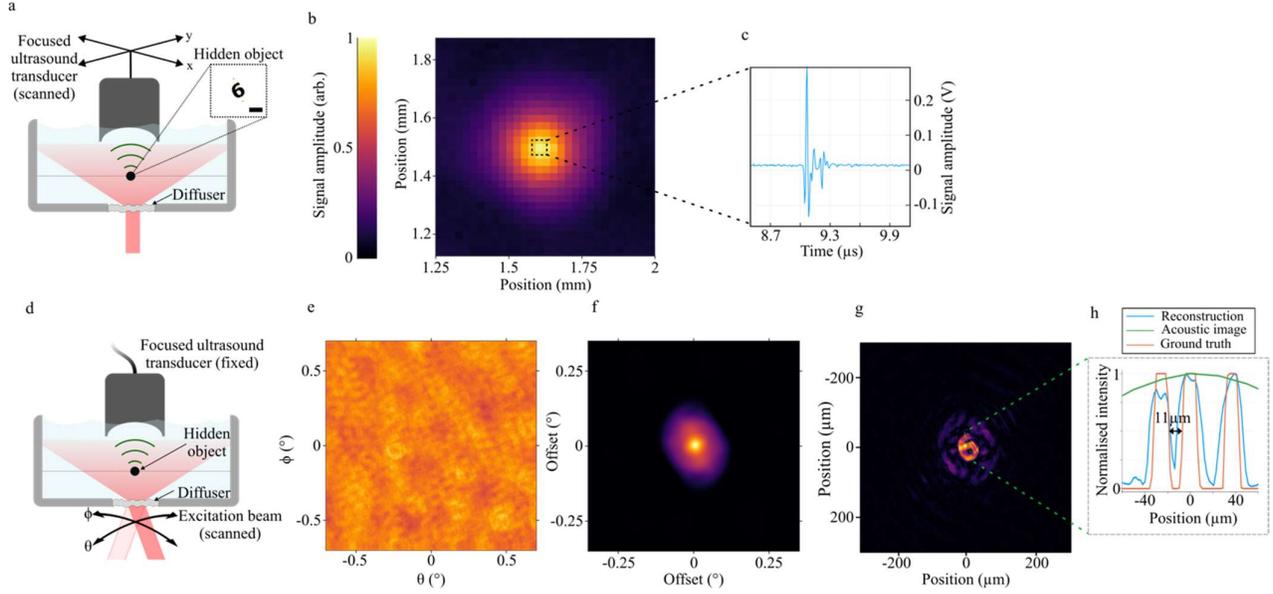

*Figure 3 – Sub-acoustic diffraction limited photoacoustic imaging through a scattering layer using speckle correlations. (a) Experimental configuration during acoustic-resolution imaging via transducer scanning. A image of the photomask design used for hidden digit is shown inset, with 50µm scale bar. (b) Acoustic resolution image of the hidden digit, with no structure visible. (c) Example time-resolved photoacoustic signal. (d) Experimental configuration during speckle correlation measurement process, with a fixed transducer location and an angle-scanned excitation beam. (e) Variation in the amplitude of photoacoustic signals as a function of excitation beam angle. (f) Autocorrelation of (e), after windowing and thresholding. (g) Reconstructed image of the hidden digit, with (h) profiles through the ground truth, reconstructed, and acoustic resolution images.*

Note that the orientation of this image has been manually flipped, as the phase retrieval process can recover either a correctly orientated image of the object or an image rotated by 180°. The digit '6' is clearly visible, indicating successful image reconstruction. The photoacoustic speckle correlation imaging process has enabled recovery of sub-acoustic spatial resolution features.

To estimate the spatial resolution enhancement provided by the photoacoustic speckle correlation imaging approach, cross-sections of the reconstructed image were compared to a ground truth reference of the mask and the acoustic resolution image. These cross sections are plotted in in Figure 3h. Features separated by a distance of 11µm are resolved, indicating a resolution at least as good as 11µm. This is significantly finer (>10×) than the resolution of the acoustic resolution image, and well below the acoustic diffraction limit at 25MHz (~40µm, larger given realistic transducer numerical apertures.).

Next, the imaging target was replaced with a film mask containing many digits arranged in a grid over a large (20×20mm) area. To investigate the ability of photoacoustic speckle correlation imaging to recover images over an area larger than the effective speckle decorrelation range, a ~2.25×2.25mm region of this grid, containing 9 distinct digits, was imaged. The imaging configuration is shown in Figure 4a.

In this work, the effective speckle correlation range was limited by the angular scan range of the illumination beam, rather than the true correlation range of the diffuser which is large due its low optical thickness. The MEMS scan range was at most ±0.5°, corresponding to a speckle scan range of ±0.12mm in the plane of the photomask, significantly smaller than the ~4mm diameter illuminated region.

As before, an acoustic resolution photoacoustic image was obtained via raster scanning of the transducer, shown in Figure 4b. As expected, this yielded an image containing nine distinct objects, with the structure of the individual objects acoustically unresolved.

Next, the photoacoustic speckle correlation measurement procedure was performed for each of these 9 objects, and individual images of each reconstructed. An image of the full field of view was assembled by centroid-aligning each reconstruction to its corresponding acoustic-resolution subregion and summing the resulting registered patches.

The complete mosaicked image is shown in Figure 4c. For clarity, an expanded view of each of the reconstructed sub-images is shown in Figure 4d, which shows reconstruction of the digits 0 through 8. Each of these images resembles the corresponding digit in the ground truth, indicating that each digit has been successfully reconstructed at significantly improved spatial resolution over the acoustic-resolution image. These results demonstrate that photoacoustic speckle correlation imaging is capable of high-resolution imaging over a field of view larger than the effective speckle decorrelation range.

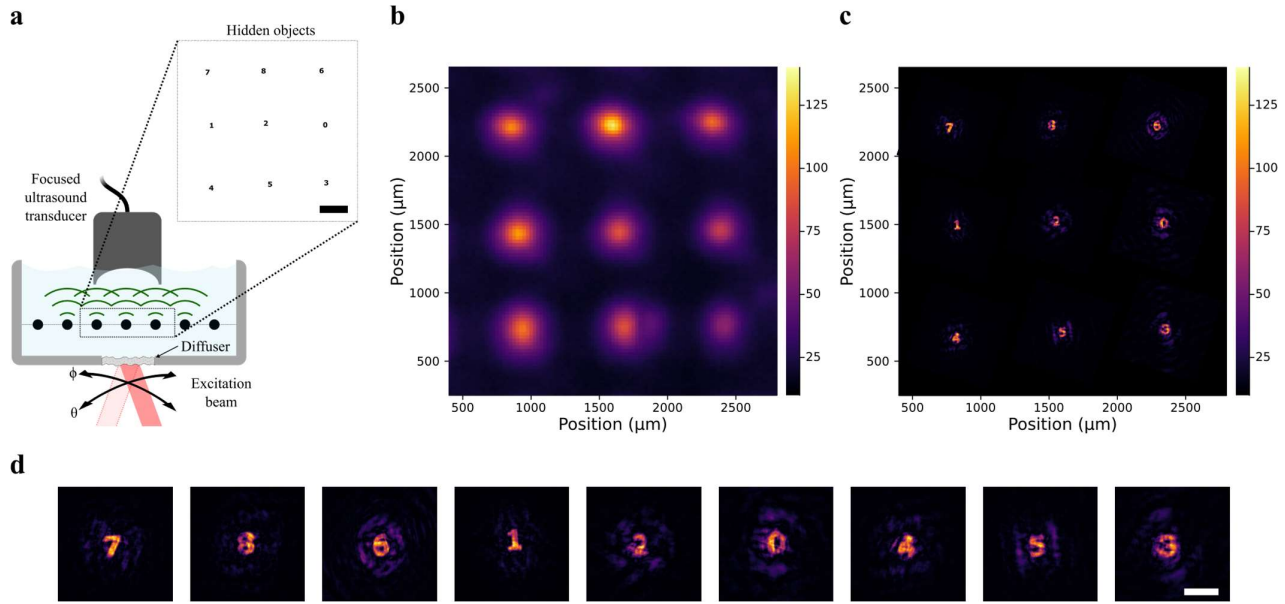

*Figure 4 – Optical resolution imaging of multiple hidden objects over an area beyond the effective memory via photoacoustic speckle correlation imaging. (a) Simplified schematic of the imaging configuration. The focused transducer was scanned in the first step of the imaging process to form an acoustic resolution image, followed by speckle correlation measurements in which the ultrasound transducer was fixed at a single position and the excitation beam angle was scanned. The photomask design used for the digit array is shown inset (500μm scale bar). (b) Acoustic resolution image of the hidden digit array. (c) Mosaicked image of the digit array at optical resolution via photoacoustic speckle correlation imaging. (d) Individual images of each reconstructed digit (100μm scale bar).*

## DISCUSSION

In this work a new photoacoustic imaging approach has been introduced. This method uses speckle patterns produced under different illumination conditions as the photoacoustic excitation, creating a sequence of distinct structured illuminations. Correlations between these patterns are then exploited to reconstruct images of optically absorbing objects through highly scattering layers with a spatial resolution exceeding the acoustic diffraction limit. This capability could provide access to microscale absorption-based structure in the multiple-scattering regime, bringing the rich multispectral and endogenous contrast central to photoacoustic imaging into a domain normally inaccessible to high-resolution methods.

In this study the spatial resolution achieved was ≤11μm. This is significantly lower than the acoustic diffraction limit for the presented system (≈125um), and even greater resolution improvements should be possible. The resolution of speckle correlation imaging approaches like the one exploited in this photoacoustic technique is expected to be determined by the size of the optical speckles illuminating the object[10,15]. This will be approximately equal to the Abbe limit of the scattered light ($\lambda/2NA$), where NA is the scattering numerical aperture[25]. In this work, with the 220 grit diffuser, the speckles grain size is expected to be <5μm, consistent with the resolution achieved. Under conditions where the NA approaches 1, enable sub-micron spatial resolution may be achievable using photoacoustic speckle correlation imaging.

Photoacoustic speckle correlation imaging enables high-resolution mapping of optical absorption. Unlike purely optical speckle correlation methods, it is also far less constrained by the usually strict requirements on extent of the speckle decorrelation range. The typical requirement for optical speckle correlation techniques is that the speckle decorrelation range is at least as large as the width of all structures contributing to measured signals. Unless the structures of interest are small and well isolated, this poses a significant challenge, as multiple scattering spreads the illumination over millimetre- to centimetre-scale regions, allowing distant structures to contribute to measured signals and demanding impractically large decorrelation ranges, which are usually significantly sub-millimetre in extent[17]. Using a photoacoustic speckle correlation approach, this requirement is relaxed so that the speckle decorrelation range need only be larger than the size of an acoustically resolvable patch[26]. With detection in the tens-of-MHz range, these acoustically resolvable patches are on the order of tens to hundreds of micrometres, compatible with more realistic decorrelation ranges.

The size of the speckle decorrelation range remains an important consideration in the application of photoacoustic speckle correlation methods due to the need to exceed the ultrasonic resolution. For methods using angular-scattering invariance for speckle translation, the decorrelation range decreases inversely with the thickness of the scattering layer, and increases with the separation between that layer and the imaging plane[13,17]. It is also substantially larger in anisotropic scattering materials, including biological tissue[17]. Consequently, when imaging objects located some distance behind a thin anisotropic scattering layer, it may be relatively straightforward to achieve acoustic resolution sufficient for photoacoustic speckle correlation imaging. For example, a recent study estimated the speckle decorrelation range at a distance of 680 μm behind a 200 μm thick layer of fixed brain tissue to be 75μm[27], within the resolution achievable by photoacoustic imaging systems[28-31].

Under more challenging conditions, such as imaging close to thicker scattering layers or within anisotropic materials, the attainable speckle decorrelation range may be substantially smaller than in the regime described above. Improving the acoustic resolution could provide one route to address this challenge, and specialist high-frequency ultrasound detectors, which can achieve spatial resolutions as small as 20 μm[31], may help accommodate such reduced decorrelation ranges. Further relaxation of decorrelation-range requirements could also be possible using more sophisticate reconstruction approaches[26,32]. Together, these strategies may provide a route to optical-resolution imaging in more demanding geometries, including situations where the structures of interest lie inside tissue.

Taking advantage of these features of photoacoustic speckle correlation imaging will require accelerating the image acquisition. In this proof-of-concept study, the acquisition time was approximately 35 minutes per image, due to a combination of the large number of angles sampled (256×256) and repeats for signal averaging. The image acquisition time could be significantly reduced by eliminating the need for signal averaging and reducing the range or sampling of the angular sweep. Sub-second image acquisition times should be possible in principle, with trade-offs between imaging speed and resolution in the limit.

Assuming that we must translate a speckle pattern across at least the width $M$ of the region of interest, equal to the acoustic resolution of the detection system, and that we must sample at least the desired reconstruction resolution $N$, then the minimum number measurements required for photoacoustic speckle correlation imaging will be $\gtrsim (M/N)^2$. In other words, the number of measurements required is proportional to the square of the desired improvement in resolution. For example, a 10× resolution improvement requires a minimum of $\gtrsim 100$ measurements. Using a 1kHz excitation laser like the one in this work, this would mean a minimum acquisition time of $\gtrsim 0.1$s. In dynamic living tissue even faster acquisition times may be needed, due to the scattering decorrelation induced by random tissue motions [33,34]. Under these conditions even greater acceleration may be necessary, targeting more modest resolution improvements or reducing the number of measurements needed using advanced reconstruction techniques[32].

In the present study, the acquisition time for imaging over a large field of view (Figure 4c) was limited by the need to serially scan the ultrasound detector. Parallel acquisition using an ultrasonic array would allow optical-resolution images of many acoustically resolvable regions to be recovered simultaneously, and would likely be necessary to apply photoacoustic speckle correlation imaging methods to large fields of view containing densely populated structural features.

At a broader conceptual level, the photoacoustic speckle correlation imaging approach taken here could be extended to exploit a number of speckle correlation effects other than scattering angle-invariance, which offer the potential of complementary capabilities. For instance, scattered light also exhibits a lateral shift invariance ('translational memory effect'), which combined with scattering angle invariance could be expected to offer an increased speckle decorrelation range[13]. Another possibility is to exploit the 'chromo-axial memory effect'[35], which produces axial speckle translation as illumination wavelength is tuned. This might enable optically-defined axial imaging resolution, and combined with the aforementioned lateral speckle shifts perhaps even full tomographic imaging at sub-acoustic spatial resolution. Speckle correlation effects present in multimode optical fibres[36-38] could even enable minimally invasive imaging beyond the depths at which other speckle correlation effects vanish, targeting applications like high-resolution imaging deep into living tissue.

**Conclusion**

We have introduced a photoacoustic imaging approach that leverages speckle correlations to provide optical-resolution imaging of absorbing structures through highly scattering layers. Through proof-of-concept experiments, we have demonstrated that this approach can enable sub-acoustic diffraction limited resolution. Additionally, we have shown that unlike optical speckle correlation techniques, this approach can image structures across fields of view larger than the effective speckle decorrelation range. The reliance on absorption contrast is intrinsically compatible with the rich multi-spectral and endogenous contrast mechanisms that underpin photoacoustic imaging in biological systems. Taken together, these capabilities suggest that correlation-based photoacoustic imaging promises a pathway to high-resolution absorption imaging in scattering environments that remain inaccessible to current modalities.

**Supplementary Methods**

Experimental setup

The full experimental setup used for photoacoustic speckle correlation imaging measurements is shown in figure S1. The pulsed laser source was a 1kHz repetition rate

Elforlight FLP, with colinear 1064nm and 532nm output. The 1064nm emission was filtered using a bandpass filter and directed to a beam dump. The 532nm emission was attenuated to ~75μJ and collimated with a diameter of ~4mm. A MEMS mirror (Mirrorcle A8L2.2-5000AL), controlled by a DAQ card (NI PCIe-6341), was used to vary the angle of this beam, which passed through a 4f telescope with a diameter of 2" and a magnification of 1, and illuminated a 220 grit ground glass diffuser. In this configuration the beam in the plane of the MEMS was imaged onto the surface of the diffuser, allowing the angle of the beam on the diffuser to be adjusted with minimal beam translation. This diffuser was mounted in a 3D printed water tank, which imaging targets could be placed within. Photoacoustic emissions produced by illuminated imaging targets were detected using a Olympus V324-SU focused ultrasound transducer. This transducer was mounted on two orthogonal motorised translation stages (Thorlabs MTS50/M-Z8) such that it could be scanned in a plane normal to the optic axis of the system. The signal from this transducer was first amplified by a custom low-noise amplifier, and then recorded using an oscilloscope (Picoscope 5443D). In all experiments the sampling rate of the oscilloscope was set to 125MHz. Trigger channels from the excitation laser, MEMS and translation stages were connected to the picoscope to allow for experimental synchronisation.

During acoustically resolution measurements, images were acquired by sequentially stepping through US transducer positions using the translation stages, and acquiring an acoustic waveform at each position. During photoacoustic speckle correlation measurements, angle-resolved acoustic measurements were acquired line-by-line, by applying a voltage ramp to one axis of the MEMS mirror, mapping angle onto signal acquisition time, and then repeating the process for another angle along the 'slow' scan axis.

Reconstruction

The basic principles of reconstruction process have been described in the main text. In overview, reconstruction involves (i) extracting photoacoustic signal amplitudes as a function of illumination angle; (ii) computing the autocorrelation of this signal and (iii) applying a windowing and thresholding operation to estimate the autocorrelation of the hidden object; and (iv) inverting this autocorrelation using phase retrieval to recover an optical-resolution image.

Photoacoustic signal amplitudes were estimated by integrating the band-limited Fourier magnitude of the relevant signal within a 0.24μs duration window. Once the autocorrelation was calculated, a background subtraction/threshold was applied by subtracting the value equal to the 30$^{th}$ percentile of the autocorrelation elements and applying a positivity constraint. The autocorrelation was then windowed by multiplication with a radial 4$^{th}$-order Butterworth filter, with a normalised cutoff value of 0.3. This corresponds to a maximum support constraint of approximately 175μm. Inversion of the autocorrelation was then performed using a continuous hybrid input-output algorithm[24], initialised with a uniform random array. As this algorithm could fail to converge on a successful image reconstruction, 100 repeats of the phase retrieval step of the reconstruction were run with different random initialisations. The image with the lowest fourier-spectrum mean-square-error was then taken as the final reconstructed image, unless this reconstruction had clearly failed in which case the image with the second or third lowest error was used instead.

In addition to these principal steps, several additions to the reconstruction algorithm were needed to compensate for practical experimental conditions.

First, while the diameter of speckle pattern in the object plane was much larger than the range over which this speckle pattern was scanned, there was still some average-intensity fall-off due to the overall beam envelope, which could interfere with reconstructions. To correct for this, the measured photoacoustic signal amplitudes were divided by a 2nd order 2D polynomial fit, which compensated for slow variation in signal amplitude due to beam fall-off.

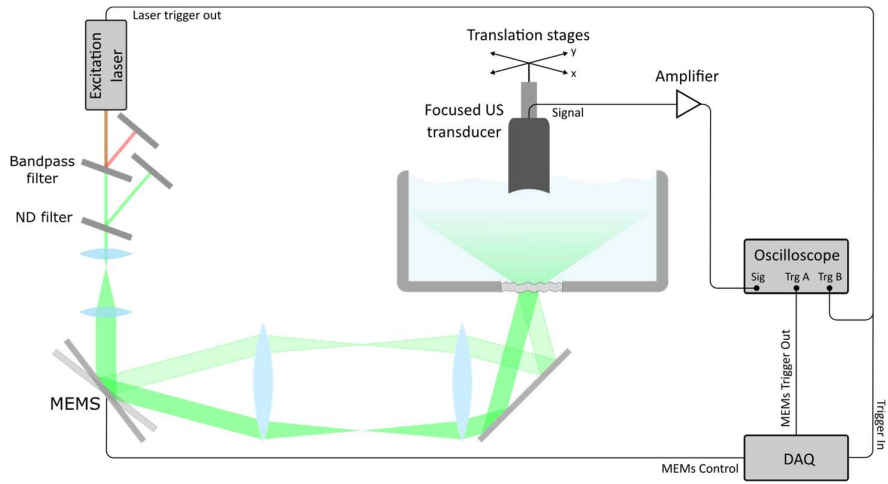

*Figure S1 – Full experimental setup.*

Second, uncorrelated measurement noise led to a strong central peak in the autocorrelation. This noise peak was removed by replacing the zero-shift autocorrelation value with the mean of the adjacent elements.

Image Orientation

One of the limitations of the technique as implemented in this work is orientation ambiguity. As discussed in the main text, it was sometimes necessary to manually re-orient reconstructed images, as the reconstruction process could recover either a correctly oriented image or an image rotated by 180°. This is a result of the fact that calculating the object autocorrelation discards absolute orientation information, such that both the object and its 180°-rotated version yield identical autocorrelations and are indistinguishable during phase retrieval.

One way to resolve ambiguity would be using image registration of adjacent or overlapping regions when mosaicking large field of view. Another method, which could work even for isolated objects without neighbours for joint registration, would be to use raw measurement data. This measurement is convolution of speckle and object – i.e. many overlapping copies of object, all with correct orientation. An area for future investigation would be automatically recovering this information as part of reconstruction process.